\documentstyle[12pt]{article}

\begin{document}

\title{Relativistic photoemission theory for
       general nonlocal potentials}

\author{C.~Meyer$^{a}$, M.~Potthoff$^{b}$, W.~Nolting$^{b}$,
        G.~Borstel$^{a}$ and J.~Braun$^{a}$}

\maketitle
\begin{center}
$^{a}$ Universit\"at Osnabr\"uck, Fachbereich Physik,
D-49069 Osnabr\"uck, Germany
\end{center}

\begin{center}
$^{b}$ Humboldt-Universit\"at zu Berlin, Institut f\"ur Physik,
D-10115 Berlin, Germany
\end{center}

\Large
{\bf Abstract}

\vspace*{0.2cm} \normalsize An improved formulation of the
one-step model of photoemission from crystal surfaces is proposed
which overcomes different limitations of the original theory.
Considering the results of an electronic-structure calculation,
the electronic (one-particle) potential and the (many-body)
self-energy, as given quantities, we derive explicit expressions
for the dipole transition-matrix elements. The theory is
formulated within a spin-polarized, relativistic framework for
general nonspherical and space-filling one-particle potentials and
general nonlocal, complex and energy-dependent self-energies. It
applies to semi-infinite lattices with perfect lateral
translational invariance and arbitrary number of atoms per unit
cell. \vspace*{0.75cm}


\newpage
\section{Introduction}
\label{sec:intro}

The spectrum of one-particle excitations of a metallic system of
correlated electrons in a solid is a fundamental question in
condensed-matter physics.
Experimentally, the interesting valence-band region around the
Fermi energy is accessible by means of ultraviolet photoemission
spectroscopy (PES) \cite{pes} and inverse
photoemission spectroscopy (IPE) \cite{ipe}.

The theoretical understanding of the excitation spectrum poses a
long-standing and not yet generally solved problem.
Within the independent-electron approximation the
spectrum is simply given in terms of the one-particle (Hartree-Fock)
eigenenergies of the Hamiltonian. Analogously, it is widely
accepted to interpret a measured photoemission spectrum by
referring to the results of band-structure calculations that are
based on density functional theory (DFT) and the local density
approximation (LDA) \cite{lda,JG89}.
Despite a sometimes convincing success in practice
\cite{SK95,Bar92,Bra96}, such an interpretation is questionable since
there is actually no known correspondence between the Kohn-Sham
eigenenergies and the one-particle excitations of the system
\cite{JG89,Bor85}.
For an in principle correct description of the excitation energies,
the local LDA exchange-correlation potential has to be supplemented
by the nonlocal, complex and energy-dependent self-energy which
leads to the Dyson equation \cite{JG89,AGD}
instead of the Schr\"odinger-type equation in
the Kohn-Sham scheme.

Provided that the self-energy is known, one can deduce a PES/IPE
``raw spectrum'' from the solution of the Dyson equation. To achieve
a reliable interpretation of experiments, however, it is inevitable
to deal with so-called ``secondary effects'' which considerably
modify and distort the raw spectrum. Above all, the wave-vector
and energy dependence of the transition-matrix elements has to be
accounted for. These dependencies are known to be important and
actually cannot be neglected. They result from strong
multiple-scattering processes which dominate the electron dynamics
in the low-energy regime of typically 1-100~eV \cite{Pen74}.
The transition-matrix elements also include the effects of selection
rules which are not accounted for in the raw spectrum.
Strictly speaking, it can be stated that the
main task of a theory of photoemission is to close the gap between
the raw spectrum obtained by (many-body) electronic-structure
calculations and the experiment.

It turns out, however, that the calculation of the
transition-matrix elements is by no means a trivial task.
Probably, the most successful theoretical approach is the
so-called one-step model of photoemission as originally proposed
by Pendry and co-workers \cite{Pen74,Pen76,HPT80}. A review on the
recent developments and refinements \cite{dev} of the approach can
be found in Ref.\ \cite{Bra96}. The main idea of the one-step
model is to describe the actual excitation process, the transport
of the photoelectron to the crystal surface as well as the escape
into the vacuum \cite{BS64} as a single quantum-mechanically
coherent process including all multiple-scattering events.

The main disadvantage of the conventional formulation of the one-step
model consists in the fact that it is intrinsically based on a
{\em local} potential. This is sufficient if the calculation
starts from the self-consistent LDA potential. The inclusion of the
{\em nonlocal} self-energy which is needed for an in principle
correct description of the one-particle excitations, however, is
not possible. In a preceding study \cite{PLNB97} we were able to
show that one can overcome this difficulty and that the nonlocal
self-energy term can be included within an alternative formulation
of the one-step model.

The purpose of the present paper is to generalize the results of
Ref.\ \cite{PLNB97} in different respects:
(i) The two-component formalism is replaced by a four-component,
relativistic framework necessary for the study of high-$Z$
materials or for spin-polarized photoemission from nonmagnetic
samples excited by circularly polarized radiation, for example.
(ii) For ferromagnetic systems, the exchange splitting and the
spin-orbit splitting are treated on equal footing. This allows to
investigate e.~g.\ magnetic dichroic effects.
(iii) The theory is no longer based on the muffin-tin approximation
for the input LDA potential.
Instead, the general case of space-filling,
nonspherical potentials is considered which becomes important
for more open crystal structures.
(iv) To cover the case of complex geometries (ordered compounds,
multilayers),
we also generalize to more than one atom per unit cell.
(v) The Korringa-Kohn-Rostocker (KKR) multiple-scattering formalism
\cite{KKR,KKR1} is used for both, the final as well as
the initial states. This internal consistency of the formalism
represents another improvement compared with Ref.\ \cite{PLNB97}
where a muffin-tin-orbitals basis was used with respect to the
initial states.

As in Ref.\ \cite{PLNB97} we keep the basic structure of the
one-step model. Starting from Pendry's formula for the photocurrent
\cite{Pen76}, analytical expressions for the matrix elements
are derived referring to a crystal surface with perfect lateral
translational invariance. For the solution of the (atomic) Dirac
equation we use the phase-functional ansatz of Calogero generalized
to the nonspherical case \cite{calogero67,gonis92}. The initial
state is treated within the relativistic version of KKR theory for
space-filling potentials \cite{KKR2,bdk96} but adapted for a slab
geometry, and the final state is constructed using a full-potential
relativistic layer-KKR method \cite{fed81,gbb}. The new formalism
is developed up to the point where the numerical evaluation has to
start.

The LDA potential and the self-energy are assumed to be given (input)
quantities that must be obtained from a
preceding electronic-structure
calculation. The related many-body problem is beyond the scope of the
present paper. In particular, one must be aware of a possible double
counting of interactions, once on a mean-field level in the LDA and
once explicitly in the self-energy.
While there are pragmatic ways to
circumvent this problem \cite{SAS92,Nolgd}, it generally remains
unsolved.
Whether or not the self-energy is calculated within a relativistic
framework is not decisive for the applicability of the presented
approach. A two-component, nonrelativistic self-energy as it is
usually provided (see e.~g.\ \cite{SAS92,Nolgd,AG95,NVF95})
can be used as an input quantity in the relativistic
formulation.

\section{One-step model of photoemission}
\label{sec:osm}

Photoemission (PES) and inverse photoemission (IPE) are complemental
spectroscopies. We concentrate on PES in the following since IPE can
simply be treated analogously by taking into account geometrical
factors that regard the respective experimental setups \cite{Pen80}.
We start our considerations by a discussion of Pendry's formula for
the photocurrent which defines the one-step model of PES
\cite{Pen76}:
\begin{equation}
  I^{\rm PES} \propto
  {\rm Im}~
  \langle \epsilon_f, {\bf k}_{\|} |
  G_{2}^+ \Delta G_{1}^+ \Delta^\dagger G^-_{2} |
  \epsilon_f, {\bf k}_{||} \rangle \: .
\label{eq:pendry}
\end{equation}

The expression can be derived from Fermi's golden rule for the
transition probability per unit time \cite{Bor85}. Consequently,
$I^{\rm PES}$ denotes the elastic part of the photocurrent.
Vertex renormalizations are neglected. This excludes
inelastic energy losses and corresponding quantum-mechanical
interference terms \cite{Bor85,Pen76,CLRRSJ73}. Furthermore,
the interaction of the outgoing photoelectron with the rest
system is not taken into account. This ``sudden approximation''
is expected to be justified for not too small photon energies.

We consider an energy-, angle- and spin-resolved photoemission
experiment. The state of the photoelectron at the detector
is written as $|\epsilon_f, {\bf k}_{\|} \rangle$, where
${\bf k}_{\|}$ is the component of the wave vector parallel
to the surface, and $\epsilon_f$ is the kinetic energy of the
photoelectron. The spin state of the photoelectron is implicit
in $|\epsilon_f, {\bf k}_{\|} \rangle$ which is understood as a
four-component Dirac spinor. The advanced Green function $G_{2}^-$
in Eq.\ (\ref{eq:pendry}) characterizes the scattering properties
of the material at the final-state energy $E_2 \equiv \epsilon_f$.
Via $|f\rangle = G^-_{2} |\epsilon_f, {\bf k}_{\|} \rangle$ all
multiple-scattering corrections are formally included. Using
standard Korringa-Kohn-Rostocker
(KKR) multiple scattering techniques
\cite{fed81}, we can calculate the final state $|f\rangle$ as a
(time-reversed) relativistic LEED state (see Sec.\ \ref{sec:fin}).

As far as concerns the final state, many-body effects are included
only phenomenologically in the LEED calculation, i.~e.\ by using a
parametrized, weakly energy-dependent and complex inner potential
$V_0(E_2)=V_{0{\rm r}}(E_2)+iV_{0{\rm i}}(E_2)$
as usual \cite{Pen74}.
This generalized inner potential also includes the (imaginary)
optical potential, which takes into account inelastic corrections
to the elastic photocurrent \cite{Bor85} as well as the actual
(real) inner potential, which serves as a reference energy inside
the solid with respect to the vacuum level \cite{HPM+95}. Due to
the finite imaginary part $i V_{0{\rm i}}(E_2)$, the flux of
elastically scattered electrons is permanently reduced, and thus
the amplitude of the high-energy wave field $|f\rangle$ can be
neglected beyond a finite distance from the surface. It is thus
sufficient to restrict oneself to a slab geometry in a practical
computation.

$\Delta$ in Eq.\ (\ref{eq:pendry}) is the dipole operator in the
electric dipole approximation which is well justified in the visible
and ultraviolet spectral range. It mediates the coupling of the
high-energy final state with the low-energy initial states. For
the relativistic, possibly ferromagnetic case and for general
space-filling potentials, a convenient form of the dipole operator
is given in Sec.\ \ref{sec:dip}.

The ``low-energy'' propagator $G_{1}^+$ in Eq.\ (\ref{eq:pendry}),
i.~e.\ the one-electron retarded Green function for the initial state
in the operator representation, yields the ``raw spectrum''. It is
directly related to the ``bare'' photocurrent and thereby represents
the central physical quantity within the one-step model.
$G_{1}^+ \equiv G_{1}^+(E_1)$
is to be evaluated at the initial-state energy $E_1 \equiv \epsilon_f
-\omega-\mu_0$, where $\omega$ is the photon energy ($\mu_0$ stands
for the chemical potential).

In the framework of the conventional one-step model of Pendry and
co-workers \cite{Pen74,Pen76,HPT80} the initial-state Green function
$G_{1}^+$ is determined for $\uparrow , \downarrow$ electrons
moving in an (effective) one-particle potential
$V^{\uparrow \downarrow}_{\rm LDA}({\bf r})$
provided by DFT-LDA. As usual,
$V^{\uparrow \downarrow}_{\rm LDA}({\bf r})$
consists of the external core potential, the Hartree
contribution as well as the exchange-correlation potential.
In the relativistic generalization of DFT \cite{raja73,rama83}
one has to consider the (one-particle) Dirac Hamiltonian
($\hbar = m = e = 1, c=137.036$):
\begin{equation}
   h_{\rm LDA}({\bf r}) = - i c \mbox{\boldmath$\alpha$}
   \mbox{\boldmath$\nabla$} +
   \beta c^{2} - c^2
   + V_{\rm LDA}({\bf r}) +
   \beta \mbox{\boldmath $\sigma$}
   {\bf B}_{\rm LDA} ({\bf r})~,
\label{eq:ldaham}
\end{equation}
where $V_{\rm LDA}({\bf r})$ denotes the (effective)
spin-independent potential, and ${\bf B}_{\rm LDA} ({\bf r})$ is the
(effective) magnetic field. They are given as \cite{strange89}:
\begin{equation}
  V_{\rm LDA}({\bf r}) =
  \frac{1}{2} (V^{\uparrow}_{\rm LDA} ({\bf r})~+~
  V^{\downarrow}_{\rm LDA} ({\bf r}))
\end{equation}
and
\begin{equation}
  {\bf B}_{\rm LDA}({\bf r}) =
  \frac{1}{2} (V^{\uparrow}_{\rm LDA} ({\bf r})~-~
  V^{\downarrow}_{\rm LDA} ({\bf r})) \: {\bf b}~.
\end{equation}
The constant unit vector ${\bf b}$ determines the spatial
direction of the (uniform) magnetization as well as the spin
quantization axis. $\beta$ denotes the usual $4 \times 4$ Dirac
matrix with the nonzero diagonal elements
${\beta}_{11}={\beta}_{22}=1$ and ${\beta}_{33}={\beta}_{44}=-1$,
and the vector $\mbox{\boldmath $\alpha$}$ is given by its
components $\alpha_k = \sigma_x \otimes \sigma_k$ ($k=x,y,z$)
in terms of the $2 \times 2$ Pauli-matrices $\sigma_k$.

Within the DFT ground-state formalism,
$V_{\rm LDA}({\bf r})$ as well as ${\bf B}_{\rm LDA}({\bf r})$
are local functions.
On the other hand, it is well known that for an in principle exact
description of the one-particle excitations one has to consider the
Dyson equation for the Green function \cite{JG89,AGD}. This
includes the nonlocal, complex and energy-dependent (retarded)
self-energy ${\Sigma}^{\uparrow \downarrow}({\bf r},{\bf r}',E)$.
As the LDA potential, the self-energy
must be assumed to be a given quantity for the photoemission theory.
Therewith, we can construct a generalized potential,
\begin{eqnarray}
  U({\bf r},{\bf r}',E) &=&
  \delta ({\bf r}-{\bf r}')~(V_{\rm LDA}({\bf r})~+~
  \beta \mbox{\boldmath $\sigma$}
  {\bf B}_{\rm LDA}({\bf r}))
\nonumber \\
&+&
  V({\bf r},{\bf r}',E)~+~
  \beta \mbox{\boldmath $\sigma$}
  {\bf B}({\bf r},{\bf r}',E)~,
\end{eqnarray}
where the nonlocal contributions $V$ and ${\bf B}$
are defined as:
\begin{equation}
V({\bf r},{\bf r}',E)~=~\frac{1}{2}({\Sigma}^{\uparrow}
({\bf r},{\bf r}',E)~+~{\Sigma}^{\downarrow}({\bf r},{\bf r}',E))
\end{equation}
and
\begin{equation}
{\bf B}({\bf r},{\bf r}',E)~=~\frac{1}{2}({\Sigma}^{\uparrow}
({\bf r},{\bf r}',E)~-~{\Sigma}^{\downarrow}({\bf r},{\bf r}',E)) \:
{\bf b}~.
\end{equation}

The initial state is described by a $4 \times 4$ Green matrix
$G_1^{+}({\bf r},{\bf r}',E_{1})$.
It can be obtained as the solution of the Dyson equation which
can be written as:
\begin{eqnarray}
  \left[ E_1 + \mu_0 -
  h_{\rm LDA}({\bf r}) \right]
  G_1^{+}({\bf r},{\bf r}',E_{1}) &-&
  \int (V({\bf r},{\bf r}'',E_{1})~+~
  \beta \mbox{\boldmath $\sigma$} {\bf B} ({\bf r},
  {\bf r}'',E_{1}))
\nonumber \\
&*&
  G_1^{+}({\bf r''},{\bf r}',E_1) d {\bf r}'' =
  \delta({\bf r}-{\bf r}')
  {{1}\mbox{\hspace{-2pt}}{\rm l}}~.
  \label{eq:eomgr}
\end{eqnarray}

A direct solution of the Dyson equation
in real-space representation turns out to be inconvenient.
Analogously to Ref.\ \cite{PLNB97}, we therefore turn to a matrix
representation and choose the eigenspinors of the LDA Hamiltonian
(\ref{eq:ldaham}) as basis states. Assuming perfect lateral
translational symmetry, the parallel component of the
wave vector is a good quantum number, and the LDA eigenvalue
problem reads:
\begin{equation}
  h_{\rm LDA} | n , {\bf q}_\| \rangle =
  {\epsilon}_{n} ({\bf q}_\|) | n , {\bf q}_\|
  \rangle \: .
\label{eq:ldaeigvalproblem}
\end{equation}
Here ${\bf q}_\|$ is a vector of the first two-dimensional
Brillouin zone, and ${\epsilon}_{n} ({\bf q}_\|)$ is the
two-dimensional band structure.
Using the eigenspinor basis, the Dyson equation for the
initial-state Green function can be written:
\begin{equation}
  \sum_{n'} \left(
  (E_1 + \mu_0 - {\epsilon}_{n} ({\bf q}_\|)) \delta_{nn'}
  - V_{nn'}(E_1 , {\bf q}_\|) - B_{nn'}(E_1 , {\bf q}_\|)
  \right)
  G^{(+)}_{n'n''}(E_1 , {\bf q}_\|) =
  \delta_{nn''}~.
\label{eq:eom}
\end{equation}
Here we have introduced the matrix representation of the nonlocal
terms $V$ and $B$:
\begin{eqnarray}
  V_{nn'}(E_1 , {\bf q}_\|) & = &
  \langle n , {\bf q}_\| | V(E_1) | n' , {\bf q}_\| \rangle \: ,
  \nonumber \\
  B_{nn'}(E_1 , {\bf q}_\|) & = &
  \langle n , {\bf q}_\| |
  \beta \mbox{\boldmath $\sigma$} {\bf B}(E_1)
  | n' , {\bf q}_\| \rangle
  \: .
\end{eqnarray}
Lateral translational symmetry requires $V$ and $B$ to be diagonal
with respect to ${\bf q}_\|$.

In the eigenspinor basis Pendry's formula reads:
\begin{equation}
  I \propto
  {\rm Im}~
  \sum_{nn'}
  M_{n} (\epsilon_f ,{\bf k}_\|)~
  G^{(+)}_{nn'}(E_1 , {\bf q}_\|)~
  M_{n'}^\ast
  ({\epsilon}_f, {\bf k}_\|)~,
\label{eq:iphototme}
\end{equation}
where
\begin{equation}
  M_{n} (\epsilon_f ,{\bf k}_\|) =
  \langle \epsilon_f, {\bf k}_\| | G_{2}^+
  \Delta | n , {\bf q}_\| \rangle
\label{eq:tmedef}
\end{equation}
is the matrix element of the dipole operator between the final state
$|f\rangle = G_{2}^{-} |{\epsilon}_f ,{\bf k}_\| \rangle$ and the LDA
eigenspinor $|n , {\bf q}_\| \rangle$. In Eq.\ (\ref{eq:iphototme})
${\bf q}_\|$ is fixed by translational symmetry: A nonzero
matrix element (\ref{eq:tmedef}) requires
${\bf q}_\| = {\bf k}_\|$ + ${\bf g}_\|$, where ${\bf g}_\|$
is a reciprocal lattice vector \cite{PLNB97}. Furthermore,
${\bf k}_\|$ is
given by the photoelectron energy at the detector and by the
emission angles. This implies that ${\bf g}_\|$ and thus
${\bf q}_\| = {\bf q}_\| ({\bf k}_\|)$ are uniquely determined
by demanding ${\bf q}_\|$ to lie within the first two-dimensional
Brillouin zone.

To work out the differences with respect to the original formulation,
let us briefly discuss the case $\Sigma \equiv 0$. Energy and
momentum conservation then implies that the final state $| f \rangle$
can couple via $\Delta$ to a single ``initial'' spinor
$| n , {\bf q}_\| \rangle$ only: Its eigenenergy is given by
${\epsilon}_{n} ({\bf q}_\|) = \epsilon_{f} - \omega$, and
${\bf q}_\| = {\bf k}_\|$ + ${\bf g}_\|$.
This implies that we can set
\begin{equation}
  G_{1}^+ = | n , {\bf q}_\| \rangle
  \frac{1}{E_1 - (\epsilon_{n}({\bf q}_\|) - \mu_0) + i 0^+}
  \langle n , {\bf q}_\| |
\end{equation}
in Eq.\ (\ref{eq:pendry}). I.~e.\ the conventional one-step model
is characterized by a one-pole structure of the low-energy (hole)
propagator. It is because of this (implicit) one-pole structure
that the original formulation does not allow to include a general
nonlocal self-energy term. (Essentially the same argument applies
if the infinitesimal $i0^+$ is replaced by a small finite imaginary
constant as is oftenly done.)

On the contrary, for the general case $\Sigma = V + B \ne 0$ we have:
\begin{equation}
  G_{1}^+ = \sum_{nn'} | n , {\bf q}_\| \rangle
  \left[\frac{{{1}\mbox{\hspace{-2pt}}{\rm l}}}
  {E_1 - (\mbox{\boldmath $\epsilon$}({\bf q}_\|) - \mu_0)
  - \mbox{\boldmath $\Sigma$}(E_1 , {\bf q}_\|)} \right]_{nn'}
  \langle n' , {\bf q}_\| | \: .
\end{equation}
The imaginary part of the self-energy causes an energy, wave-vector
and band dependent broadening of the initial state. The hole
acquires a finite lifetime (except for $E_1=0$, i.~e.\ for
$\epsilon_f-\omega = \mu_0$ where the self-energy
is Hermitian). The energy dependence of $\Sigma$ may introduce
a band-narrowing effect, an enhancement of the effective electron
mass at the Fermi edge and, for a strongly correlated system, may
give rise to satellite features in the spectrum. All this implies
that for a given photon energy $\omega$ the final state
$| f \rangle$ necessarily couples to different initial states
$|n, {\bf q}_\| \rangle$. Furthermore, since $\Sigma$
is generally nondiagonal in the band index $n$, interference terms
$n\ne n'$ have to be considered.

The number $N_n$ of possible values for the band index $n$ in
Eqs.\ (\ref{eq:eom}) and (\ref{eq:iphototme}) is given by
$N_n = N_\perp N_{\sf A} N_{\sf K}$ where $N_\perp$ is the
number of layers in the slab and $N_{\sf A}$ is the number of atoms
in the two-dimensional unit cell. $N_{\sf K}$ is determined
by the maximum order in the spin-angular momentum expansion (see
next section) that is necessary for convergence. $N_\perp$
can be assumed to be finite since the damping of the final-state
wave field implies that contributions to the photocurrent are
negligibly small beyond a finite distance from the surface. In
fact, a finite $N_\perp$ is decisive for a numerical solution of
the Dyson equation (\ref{eq:eom}) by matrix inversion. One inversion
is necessary to get one $(\epsilon_f, {\bf k}_\|)$ point in the
PES spectrum.

Formulas (\ref{eq:eom}), (\ref{eq:iphototme}) and (\ref{eq:tmedef})
generalize our approach of Ref.\ \cite{PLNB97}. The latter is
obtained in the nonrelativistic approximation. On the other hand,
in the limit
${\Sigma}^{\uparrow \downarrow}({\bf r},{\bf r}',E) \equiv 0$
the initial state is treated as in the original (relativistic)
formulation of the one-step model \cite{Bra96,dev}.

The actual and
remaining task, however, consists in the according recalculation
of the transition-matrix elements (\ref{eq:tmedef}). This is done
in the following. We thus consider the initial states
$|n , {\bf q}_\| \rangle$ (Sec.\ \ref{sec:ini}),
the final state
$G_{2}^- | \epsilon_f, {\bf k}_\| \rangle$ (Sec.\ \ref{sec:fin}),
and the dipole operator
$\Delta$ (Sec.\ \ref{sec:dip}) to derive the final expression
for the matrix elements in Sec.\ \ref{sec:tme}.

\section{The initial states}
\label{sec:ini}

For the evaluation of the matrix elements, we need a one-center
expansion of the LDA eigenstates $|n,{\bf q}_\|\rangle$.
This will finally allow for a separation of the matrix elements
into radial and angular parts. Such a one-center expansion is
available within the relativistic version of the KKR
approach for arbitrary space-filling potentials \cite{KKR1}.

We consider the slab to be built up from layers
$i_\perp = 1,...,N_\perp$ parallel to the surface.
The two-dimensional unit cells within a layer $i_\perp$ are
labeled by an index $i_\|$ and the atoms within the unit cells by
an index ${i_{\sf A}}$.
The position vector of a particular atom in the
semi-infinite lattice
$i=(i_\|,i_\perp,{i_{\sf A}})$ is then given by:
${\bf R}_{i}={\bf R}_{i_\|}+{\bf R}_{i_\perp}+{\bf R}_{{i_{\sf A}}}$,
where ${\bf R}_{i_\|}$ is a vector of the two-dimensional lattice,
${\bf R}_{i_\perp}$ denotes the local origin of layer $i_\perp$,
and ${\bf R}_{{i_{\sf A}}}$
the position vector of the ${i_{\sf A}}$-th atom
with respect to the local origin of the unit cell. The material
space is decomposed into (three-dimensional) polyhedra (atomic cells)
$\Omega_i$ with one atom $i$ at the center of each such that
the LDA potential $V_{\rm LDA}({\bf r})$ (and analogously the
effective magnetic field ${\bf B}_{\rm LDA}({\bf r})$)
can be written as a sum over cell potentials $V_i({\bf r}-{\bf R}_i)$
that vanish outside $\Omega_i$:
\begin{equation}
   V_{\rm LDA}({\bf r}) = \sum_i V_i({\bf r}-{\bf R}_i) \: .
\label{eq:decomp}
\end{equation}
A cell $\Omega_i$ is circumscribed by a bounding sphere $S_i$ of
radius $R_i$. Due to translational symmetry the potential within
the cell $\Omega_i$ or within the sphere $S_i$ only depends on
$i_\perp$ and ${i_{\sf A}}$. We have:
$V_i({\bf r}) = V_{i_\perp {i_{\sf A}}}({\bf r})$. The same notation
also applies for the effective magnetic field etc.

For the practical calculation, slabs consisting of $N_\perp$ layers
each are arranged in a super-cell geometry with a sufficiently
large distance in between. This formally restores full 
three-dimensional periodicity and thus allows the application of 
the conventional KKR method.
It can be shown \cite{bdk96} that the wave function
$\Psi^{(n)}_{{\bf q}_\|}({\bf r}) \equiv
\langle {\bf r} | n , {\bf q}_\| \rangle$
of the initial spinor $| n , {\bf q}_\| \rangle$
can be expanded in locally exact basis spinors
$\Phi_{i\kappa\mu}(E,r)=\Phi_{i_\perp{i_{\sf A}}\kappa\mu}(E,r)$:
\begin{equation}
  \Psi^{(n)}_{{\bf q}_\|}({\bf r}) =
  \sum_{\kappa \mu}
  A_{i \kappa \mu} (\epsilon_{n}({\bf q}_\|) , {\bf q}_\|)~
  \Phi_{i\kappa\mu}(\epsilon_{n}
    ({\bf q}_\|),{\bf r} - {\bf R}_i) \: .
\label{eq:infinal}
\end{equation}
The expansion converges within each cell $\Omega_i$ and for each
$\epsilon_n({\bf q}_\|)$ \cite{bdk96}. The basis spinors satisfy
\begin{equation}
   \left(
   E + i c \mbox{\boldmath $\alpha$}
   \mbox{\boldmath$\nabla$}
   - \beta c^{2} +c^2 - V_{i}({\bf r}) -
   \beta \mbox{\boldmath $\sigma$}
   {\bf B}_{i} ({\bf r})
   \right)
   \Phi_{i\kappa\mu}(E,{\bf r})
   = 0
\label{eq:diracatom}
\end{equation}
for all ${\bf r}$. They differ in their behavior close to the
center of $\Omega_i$. For $|{\bf r}|\mapsto 0$:
\begin{equation}
   \Phi_{i\kappa\mu}(E,{\bf r}) = J_{\kappa}^{\mu} (k,{\bf r}) \: ,
\label{eq:diracatom1}
\end{equation}
where
\begin{equation}
  J_{\kappa}^{\mu} (k,{\bf r}) = \left(
  \begin{array}{c}
    \chi_{\kappa}^{\mu} ({\bf \hat{r}}) \; j_{l} (k r) \\
    i \chi_{-\kappa}^{\mu} ({\bf \hat{r}})
    \left( \frac{k S_{\kappa} c}{E + 2 c^{2}} \right)
    j_{\bar{l}} (k r)
  \end{array}
  \right) \: ,
\label{eq:bessel}
\end{equation}
with the spherical Bessel function $j_{l}(kr)$
and $k$, $S_\kappa$, and $\bar{l}$ defined by
$k=\sqrt{2E+E^2/c^2}$,
$S_\kappa = \kappa/|\kappa|$, and
$\bar{l}=l-S_\kappa$.
$\kappa$ and $\mu$ are the relativistic spin-angular momentum
indices according to Rose \cite{rose61}.
The spin-angular functions
$\chi_{\kappa}^{\mu}({\bf \hat{r}}) = \sum_s C_{\kappa \mu s}
Y_l^{\mu-s}({\bf \hat{r}}) \chi_s$
are given in the usual way \cite{rose61} in terms of Clebsch-Gordan
coefficients $C_{\kappa \mu s}$, spherical harmonics $Y_l^m$ and
Pauli spinors
$\chi_{1/2} = (1,0)^\dagger$, $\chi_{-1/2} = (0,1)^\dagger$.

Eq.\ (\ref{eq:diracatom}) is a well-defined (effectively atomic)
problem which together with (\ref{eq:diracatom1}) uniquely
determines the $\Phi_{i\kappa\mu}(E,{\bf r})$. It can be solved
using the (nonspherical) phase-functional ansatz of Calogero
generalized to the relativistic case
\cite{calogero67,gonis92}. We separate radial
and angular parts and write:
\begin{eqnarray}
  {\Phi}_{i\kappa \mu}(E,{\bf r})
&=& \sum_{\kappa'\mu'}
  \left( \begin{array}{r}
     \chi_{\kappa'}^{\mu'}({\bf \hat{r}})~
     \phi_{i \kappa' \mu' \kappa \mu}^{u}(E,r) \\
    i\chi_{-\kappa'}^{\mu'}({\bf \hat{r}})~
     \phi_{i \kappa' \mu' \kappa \mu}^{l}(E,r) \\
  \end{array} \right)
\nonumber \\
& \equiv & \sum\limits_{\kappa^\prime \mu^\prime} \left(
  J_{\kappa^\prime}^{\mu^\prime} (k,{\bf r})
  C_{i\kappa' \mu' \kappa \mu}(E,r) -
  N_{\kappa^\prime}^{\mu^\prime} (k,{\bf r})
  S_{i\kappa' \mu' \kappa \mu}(E,r) \right) \: ,
\label{eq:psifin}
\end{eqnarray}
where $N_{\kappa}^{\mu} (k,{\bf r})$ is defined analogous to Eq.\
(\ref{eq:bessel}) replacing $j_{l}(kr)$ by the spherical Neumann
functions $n_{l}(kr)$. The superscripts $u$ and $l$ refer to
the upper and lower components of the four-component spinor,
respectively. Inserting into (\ref{eq:diracatom}) eventually
yields the coupled channel equations \cite{bdk96}
for the coefficient
matrices $C_i$ and $S_i$. The coupled channel equations may be
solved by outward integration from the origin to the radius of
the bounding sphere $R_i$ for any effective potential that is
less singular than $r^{-2}$ ($p \equiv k (E+2c^2) /c$):
\begin{eqnarray}
C_{i\kappa\mu\kappa'\mu'} (E,r) & = &
\delta_{\kappa\kappa'} \delta_{\mu\mu'}
-p \int_0^r r^2 dr \int_{(4 \pi )} d{\bf \hat{r}} \;
N_{\kappa}^{\mu}(k,{\bf r})^\dagger
  \left( V_i({\bf r}) + \beta \mbox{\boldmath $\sigma$}
  {\bf B}_i({\bf r}) \right)
\Phi_{i\kappa'\mu'} (E,{\bf r})
\: , \nonumber \\
S_{i\kappa\mu\kappa'\mu'} (E,r) & = &
- p \int_0^r r^2 dr \int_{(4 \pi )} d{\bf \hat{r}} \;
J_{\kappa}^{\mu}(k,{\bf r})^\dagger
  \left( V_i({\bf r}) + \beta \mbox{\boldmath $\sigma$}
  {\bf B}_i({\bf r}) \right)
\Phi_{i\kappa'\mu'} (E,{\bf r}) \: .
\label{eq:cce}
\end{eqnarray}
Expanding $V_i({\bf r})$ and ${\bf B}_i({\bf r})$ in spherical
harmonics, the angular integrations can be done analytically.
Outside the bounding sphere where
$V_i({\bf r})={\bf B}_i({\bf r})=0$, the coefficient matrices
$C_{i\kappa\mu\kappa'\mu'}(E,r)$ and
$S_{i\kappa\mu\kappa'\mu'}(E,r)$ are constant.

Using the result (\ref{eq:cce}) in Eq.\ (\ref{eq:psifin}) and
inserting into Eq.\ (\ref{eq:infinal}) yields the desired one-center
expansion of the initial states. The coefficients
$A_{i \kappa \mu} (\epsilon_{n}({\bf q}_\|),{\bf q}_\|)$
in (\ref{eq:infinal}) can be obtained from the KKR secular
equation \cite{bdk96} which for a slab geometry reads:
\begin{eqnarray}
  &&
  \sum_{i_\perp' {{i_{\sf A}}'}} \sum_{\kappa'\mu'} \Bigg(
  \delta_{i_\perp {i_\perp}'} \delta_{{i_{\sf A}} {{i_{\sf A}}'}}
  C_{{i_\perp}'{{i_{\sf A}}'} \kappa\mu \kappa'\mu'}(E,R_{i'})
  \nonumber \\ &&
  -\sum_{\kappa''\mu''}
  B_{i_\perp{i_{\sf A}}\kappa\mu,
  {i_\perp}'{{i_{\sf A}}'}\kappa''\mu''}(E,{\bf q}_\|) ~
  S_{{i_\perp}'{{i_{\sf A}}'} \kappa''\mu''\kappa'\mu'}(E,R_{i'})
  \Bigg)
  A_{{i_\perp}' {{i_{\sf A}}'} \kappa' \mu'}(E,{\bf q}_\|)
  = 0 \: .
\label{eq:secul}
\end{eqnarray}
$B$ are the usual (three-dimensional) KKR structure constants
(cf.\ Ref.\ \cite{bdk96}) for the super-cell geometry (each 
super-cell consists of $N_\perp \times N_{\sf A}$ atoms).

\section{The final state}
\label{sec:fin}

The final state $\Psi_{{\bf k}_\|}^{(f)}({\bf r}) \equiv \langle
{\bf r} | G_{2}^-|\epsilon_f, {\bf k}_\| \rangle$ is an eigenspinor
of $h_{\rm LDA}$ with eigenenergy $E_2=\epsilon_f$ and could thus be
constructed in the same way as the initial states. However, for an
appropriate description of the photoemission process we must ensure
the correct asymptotic behavior of
$\Psi_{{\bf k}_\|}^{(f)}({\bf r})$
beyond the crystal surface, i.~e. a single outgoing plane wave
characterized by $\epsilon_f$ and ${\bf k}_\|$. Furthermore, the
damping of the final state due to the imaginary part of the inner
potential $iV_{0{\rm i}}(E_2)$ must be taken into account. We thus
construct the final state within SPLEED theory considering a single
plane wave $|\epsilon_f, {\bf k}_\| \rangle$ {\em advancing} onto
the crystal surface. Using the standard layer-KKR method \cite{KKR}
generalized for the relativistic, full-potential case (cf.\ e.~g.\
Ref.\ \cite{Bra96}), we first obtain the SPLEED state
$U \Psi_{{\bf k}_\|}^{(f)}({\bf r})$. The final state is then given
as the time-reversed SPLEED state
($U=-i \sigma_y K$ is the relativistic time inversion).

The scattering properties of an atomic cell $\Omega_i$ can be
determined from the solution of the coupled channel equations
(\ref{eq:cce}) for the final-state energy ${\epsilon}_f$.
From $C_{i\kappa\mu\kappa'\mu'}(E=\epsilon_f,r=R_i)$ and
$S_{i\kappa\mu\kappa'\mu'}(E=\epsilon_f,r=R_i)$ the atomic
scattering matrix $\Gamma_{i\kappa\mu\kappa'\mu'}$ is
constructed as:
\begin{equation}
    \Gamma_{{i\kappa}{\mu}{\kappa}'{\mu}'}= \frac{1}{2}
    \left(
    \sum\limits_{{\kappa}''{\mu}''}
    U_{{i\kappa}{\mu}{\kappa}''{\mu}''}
    V^{-1}_{{i\kappa}'' {\mu}'' {\kappa}' {\mu}'}
    \; - \;
    \delta_{\kappa\kappa'}
    \delta_{\mu\mu'}
    \right) \: ,
\end{equation}
where the new coefficient matrices $U_i$ and $V_i$ are given by:
\begin{eqnarray}
 U_{{i\kappa}{\mu}{\kappa}'{\mu}'} & = &
   C_{{i\kappa}{\mu}{\kappa}'{\mu}'}(\epsilon_f,R_i)
   \;\; + \; i \;
   S_{{i\kappa}{\mu}{\kappa}'{\mu}'}(\epsilon_f,R_i) \; , \\
 V_{{i\kappa}{\mu}{\kappa}'{\mu}'} & = &
   C_{{i\kappa}{\mu}{\kappa}'{\mu}'}(\epsilon_f,R_i)
   \;\; - \; i \;
   S_{{i\kappa}{\mu}{\kappa}'{\mu}'}(\epsilon_f,R_i) \; .
\end{eqnarray}
The matrix $\Gamma_i \equiv \Gamma_{i_\perp {i_{\sf A}}}$
together with
the crystal geometry determine the scattering matrix $M_{i_\perp}$
for a single layer $i_\perp$ \cite{Bra96}:
\begin{eqnarray}
   M_{i_{\perp} {\bf gg'}}^{\tau \tau' ss'} &=&
   \delta_{{\bf gg'}}^{\tau \tau' ss'} +
   \frac{8 {\pi}^{2}}{k k_{ {\bf g} z}^{+}}
     \sum\limits_{\kappa \mu}
     \sum\limits_{{\kappa}' {\mu}'}
     \sum\limits_{{\kappa}'' {\mu}''}
     \sum\limits_{{i_{\sf A}} {{i_{\sf A}}'}}
      i^{-l}
      C_{{\kappa} {\mu} s}
      Y_{l}^{\mu -s}
      ({\bf \hat{k}}^{\tau}_{ \bf g})
      e^{-i {\bf k}_{ {\bf g}}^{\tau} {\bf R}_{{i_{\sf A}}} }
   \nonumber \\
     &*&
     {\Gamma}_{i_{\perp} {i_{\sf A}},
     {\kappa} {\mu} {\kappa}'' {\mu}''}
     (1-X)^{-1}_{i_{\perp}, {i_{\sf A}} {\kappa}''
     {\mu}'', {{i_{\sf A}}'} {\kappa}' {\mu}'}
      i^{l'}
      C_{{\kappa}' {\mu}' s'}
      Y_{l'}^{{\mu}' -s'}
      ({\bf \hat{k}}^{\tau'}_{ \bf g'})^\ast
      e^{i {\bf k}_{ {\bf g}'}^{\tau'}
      {\bf R}_{{{i_{\sf A}}'}} } \: .
\label{eq:mmat}
\end{eqnarray}
${\bf k}_{ {\bf g}}^{\pm}$ denotes the wave vector and $\tau=\pm$
defines the direction of a plane wave (incoming or outgoing) with
respect to the layer $i_\perp$. The parallel component of
${\bf k}_{ {\bf g}}^{\pm}$ differs from ${\bf k}_\|$ by a
two-dimensional reciprocal lattice vector ${\bf g}$ while
the perpendicular component is fixed by the plane wave energy
$\epsilon_f$. Multiple scattering processes within the layer
$i_\perp$ are taken into account via the $X$ matrix of layer-KKR
theory \cite{Pen74,KKR}:
\begin{eqnarray}
   X_{i_{\perp},{i_{\sf A}}\kappa\mu,{{i_{\sf A}}'}\kappa'\mu'}
   (\epsilon_f,{\bf k}_\|)
   &=&
   \sum_{\kappa''\mu''s} {\sum_{i_\|}}'
   e^{i {\bf k}_{\|}
   ({\bf R}_{{{i_{\sf A}}'}}-{\bf R}_{{i_{\sf A}}}-{\bf R}_{i_\|})}
   C_{\kappa\mu s}
\nonumber \\ & * &
   G_{l \mu -s , l'' \mu'' -s}
   ({\bf R}_{{{i_{\sf A}}'}}-{\bf R}_{{i_{\sf A}}}-{\bf R}_{i_\|})
   C_{{\kappa}'' {\mu}'' s}
   {\Gamma}_{i_\perp {{i_{\sf A}}'},\kappa''\mu''\kappa'\mu'} \: .
\end{eqnarray}
The prime on the summation over $i_\|$ indicates that the term
where ${\bf R}_{i_\|} = 0$ and at the same time
${i_{\sf A}} = {{i_{\sf A}}'}$
is omitted. An explicit expression for the lattice sum
$G_{l m,l'm'} ({\bf R}_{{{i_{\sf A}}'}} - {\bf R}_{{i_{\sf A}}}
-{\bf R}_{i_\|})$ can be found in Ref.\ \cite{Pen74}. 

Therewith, the scattering properties of all layers are known. We
consider a plane wave $|\epsilon_f, {\bf k}_\| \rangle$ advancing
onto the crystal from the vacuum side. From the scattering matrices
(\ref{eq:mmat}) the coefficients $u^\tau_{i_{\perp} {\bf g} s}$ of
an expansion into plane waves $|{\bf g}, \tau, s \rangle$ in front of
all layers $i_\perp$ can easily be found using standard recursive
layer-by-layer schemes \cite{Pen74,fed81}. Time reversal then yields
the final-state wave function $\Psi_{{\bf k}_\|}^{(f)}({\bf r})$.

Within each atomic cell $\Omega_i$ the final state may be expanded,
\begin{equation}
   \Psi_{{\bf k}_\|}^{(f)}({\bf r}) = \sum\limits_{\kappa\mu}
   A_{i\kappa\mu}(\epsilon_f, {\bf k}_\|)^\ast~
   \Phi_{i\kappa\mu}^{(T)}(\epsilon_f,{\bf r} - {\bf R}_i) \: ,
\label{eq:finstat}
\end{equation}
in the locally exact basis of time-reversed (``T'') phase functions:
\begin{equation}
  \Phi^{(T)}_{i\kappa\mu}(E,{\bf r})
  = \sum_{\kappa'\mu'}
  \left( \begin{array}{r}
     \chi_{\kappa'}^{\mu' (T)}({\bf \hat{r}})~
     \phi_{i \kappa' \mu' \kappa \mu}^{u}(E,r)^\ast \\
   -i\chi_{-\kappa'}^{\mu' (T)}({\bf \hat{r}})~
     \phi_{i \kappa' \mu' \kappa \mu}^{l}(E,r)^\ast \\
  \end{array} \right)
  \: .
\end{equation}
Here $\chi_{\kappa}^{\mu (T)} ({\bf \hat{r}})$ is a
time-reversed spin-angular function:
\begin{equation}
  \chi_{\kappa}^{\mu (T)} ({\bf \hat{r}}) \;
  \equiv U \chi_{\kappa}^{\mu} ({\bf \hat{r}}) \;
  = \sum\limits_{s} (-2s) C_{{\kappa} {\mu} -s}
  Y_{l}^{\mu + s}
  ({\bf \hat{r}})^\ast \chi_{s} \: .
\end{equation}
The expansion coefficients in (\ref{eq:finstat}) are derived within
SPLEED theory as usual. Starting from the bare coefficients,
\begin{eqnarray}
 A_{i_{\perp}{i_{\sf A}}\kappa\mu}^{(0)} &=&
 \sum_{\kappa'\mu'} \sum_{{\bf g}s}
 4\pi i^{l'}(-2s)(-)^{\mu'-s} C_{{\kappa'}{\mu'}s}
 ~V^{-1}_{i_{\perp} {i_{\sf A}} , \kappa\mu \kappa'\mu'}~*~
\nonumber \\
&&
 \Big[
 u^{+}_{i_{\perp} {\bf g}s}~
 Y^{s-{\mu'}}_{l'}(\widehat{{\bf k}^{+}_{\bf g}})~
 e^{i{\bf k}^{+}_{\bf g} \cdot {\bf R}_{{i_{\sf A}}}}+
 u^{-}_{i_{\perp} {\bf g}s}~
 Y^{s-{\mu'}}_l(\widehat{{\bf k}^{-}_{\bf g}})~
 e^{i{\bf k}^{-}_{\bf g} \cdot {\bf R}_{{i_{\sf A}}}} \Big] \: ,
\end{eqnarray}
and correcting for intra-layer multiple scattering, we obtain:
\begin{equation}
  A_{i_{\perp} {i_{\sf A}} {\kappa} {\mu}} =
  \sum\limits_{{\kappa}' {\mu}'}
  \sum\limits_{{\kappa}'' {\mu}''}
  \sum\limits_{{\kappa}''' {\mu}'''}
  \sum\limits_{{{i_{\sf A}}'}}
  V^{-1}_{i_{\perp} {i_{\sf A}}, {\kappa} {\mu} {\kappa}' {\mu}'}
  (1-X)^{-1}_{i_{\perp},{i_{\sf A}}\kappa'\mu' ,
    {{i_{\sf A}}'}\kappa''\mu''}
  V_{i_{\perp} {i_{\sf A}}', \kappa'' \mu'' \kappa''' \mu'''}
  A_{i_{\perp} {i_{\sf A}}' {\kappa}''' {\mu}'''}^{(0)} \: .
\end{equation}

The one-center expansion (\ref{eq:finstat}) is formally very
similar to the one-center expansion (\ref{eq:infinal}) of the
initial states. The main difference, however, is the calculation
of the expansion coefficients. For the final state they are
obtained from (full-potential) SPLEED theory
instead of solving the KKR secular equation. Thereby, we
account for the final-state damping effects and the correct
asymptotics corresponding to the experimental situation.

\section{The dipole operator}
\label{sec:dip}

In the relativistic theory the dipole interaction of an electron
with the electromagnetic field is given by the dipole operator
$\Delta = - {\mbox{\boldmath$\alpha$}} {\bf A}_0$
where ${\bf A}_{0}$
is the spatially constant vector potential inside the crystal.
In a matrix element $\langle \Psi_f | \Delta | \Psi_i \rangle$
between eigenspinors $| \Psi_f \rangle$ and $| \Psi_i \rangle$
of the Dirac Hamiltonian (\ref{eq:ldaham}) with energies $E_f$
and $E_i$, respectively, $\Delta$ can be written as:
\begin{equation}
\Delta({\bf r}) = E_{fi}
\left(
   {\bf A}_0 \mbox{\boldmath$\nabla$}
+  \frac{i\omega}{c} \mbox{\boldmath${\alpha}$} {\bf A}_0
\right)
V_{\rm LDA}({\bf r})
+ E_{fi}
\left(
   {\bf A}_0 \mbox{\boldmath$\nabla$}
\right)
\beta \mbox{\boldmath $\sigma$} {\bf B}_{\rm LDA}({\bf r})
+ E_{fi} \frac{\omega}{c} \beta {\bf A}_0 \times
\mbox{\boldmath $\sigma$} {\bf B}_{\rm LDA}({\bf r})
\: ,
\label{eq:diprel}
\end{equation}
with $E_{fi}=-2ic/[(E_f+c^2)^2-(E_i+c^2)^2]$. The expression is
derived by making use of commutator and anticommutator rules
analogously to the nonrelativistic case in Ref.\ \cite{PLNB97}.
Using the decomposition of the LDA potential and the effective
magnetic field (\ref{eq:decomp}), we immediately get:
\begin{equation}
  \Delta({\bf r}) = \sum_i \Delta_i({\bf r})  \: ,
\label{eq:deldec}
\end{equation}
where $\Delta_i({\bf r})$ vanishes outside $\Omega_i$. In the
following we calculate the dipole operator for a given atomic
cell. Distinguishing between contributions due to
$V_i({\bf r})$ and due to ${\bf B}_i({\bf r})$, the $4 \times 4$
matrix in (\ref{eq:diprel}) can be written more explicitly as:
\begin{equation}
\Delta_i({\bf r})  =
E_{fi}
\left( \left(
\begin{array}{cc}
  \Delta^{(uu)}_{V,i}({\bf r}) &
  \Delta^{(ul)}_{V,i}({\bf r}) \\
  \Delta^{(lu)}_{V,i}({\bf r}) &
  \Delta^{(ll)}_{V,i}({\bf r})
\end{array} \right) +
\left(
\begin{array}{cc}
  \Delta^{(uu)}_{B,i}({\bf r}) &
  \Delta^{(ul)}_{B,i}({\bf r}) \\
  \Delta^{(lu)}_{B,i}({\bf r}) &
  \Delta^{(ll)}_{B,i}({\bf r})
\end{array}
\right) \right) \: ,
\end{equation}
with the following $2\times 2$ matrices:
\begin{eqnarray}
\begin{array} {ll}
\Delta_{V,i}^{(uu)} ({\bf r}) =
    {{\bf A}_0{\mbox{\boldmath$\nabla$}} V_i({\bf r})}
{{1}\mbox{\hspace{-2pt}}{\rm l}}_{(2 \times 2)}
    {\mbox{\hspace{5mm}}}
&
\Delta_{V,i}^{(ul)} ({\bf r}) =
{\frac{i \omega}{c} \mbox{\boldmath $\sigma$}
{\bf A}_0 V_i({\bf r})} \\
\Delta_{V,i}^{(lu)} ({\bf r}) = \Delta_{V,i}^{(ul)}({\bf r})
    {\mbox{\hspace{5mm}}}
&
\Delta_{V,i}^{(ll)} ({\bf r}) = \Delta_{V,i}^{(uu)} ({\bf r})
 \\ & \\
\Delta_{B,i}^{(uu)} ({\bf r}) = ({\bf A}_0 {\mbox{\boldmath$\nabla$})
(\mbox {\boldmath $\sigma$} {\bf B}_i({\bf r})})
 - \frac{\omega}{c} \mbox{\boldmath $\sigma$} {\bf A}_0
   \times {\bf B}_i({\bf r})
   {\mbox{\hspace{5mm}}}
&
\Delta_{B,i}^{(ul)} ({\bf r})= 0
 \\
\Delta_{B,i}^{(lu)} ({\bf r})= 0  {\mbox{\hspace{5mm}}}
&
\Delta_{B,i}^{(ll)} ({\bf r})= - \Delta_{B,i}^{(uu)} ({\bf r}) \: .
\end{array}
\label{eq:delta1}
\end{eqnarray}

As for the initial states and the final state we need a separation
into radial and angular parts. For this purpose the cell potential
$V_i({\bf r})$ inside the bounding sphere is expanded into spherical
harmonics,
$V_i({\bf r}) = \sum_{lm} V_{ilm} (r) Y^{m}_{l} (\hat{\bf r})$
(and ${\bf B}_i({\bf r}) = {\bf b} \sum_{lm} B_{ilm} (r) Y^{m}_{l}
(\hat{\bf r})$). The expansion has to be done in a way that
guarantees
$V_{\rm LDA}({\bf r})={\bf B}_{\rm LDA}({\bf r})=0$ in the segments
between the bounding sphere and the atomic cell $(S_i - \Omega_i)$.
For the two terms in (\ref{eq:delta1}) involving the gradient the
expansion needs to be considered in detail:
\begin{eqnarray}
{{\bf A}_0{\mbox{\boldmath$\nabla$}} V_i({\bf r})} & = &
\sum\limits_{a=1}^2 \sum \limits_{l m}
V^{(a)}_{i l m}(r) W^{(a)}_{l m} (\hat{{\bf r}}) \; \; ,
\\
({{\bf A}_0 {\mbox{\boldmath$\nabla$}})
(\mbox{\boldmath $\sigma$} {\bf B}_i({\bf r})})
& = &
\sum\limits_{a=1}^2 \sum\limits_{l m}
B^{(a)}_{i l m}(r)
(\mbox{\boldmath $\sigma$} \cdot {\bf b} )
W^{(a)}_{l m} (\hat{{\bf r}}) \; \; .
\end{eqnarray}
A straightforward calculation yields
\begin{eqnarray}
  V^{(1)}_{ilm}(r) & = & \left( \frac{\partial}{\partial r} +
  \frac{l+1}{r} \right)
  V_{ilm}(r) \; , \hspace{5mm}
  V^{(2)}_{ilm}(r) \: = \: \left( \frac{\partial}{\partial r} -
  \frac{l}{r} \right)
  V_{ilm}(r) \: , \\
  B^{(1)}_{ilm}(r) & = & \left( \frac{\partial}{\partial r} +
  \frac{l+1}{r} \right)
  B_{ilm}(r) \; , \hspace{5mm}
  B^{(2)}_{ilm}(r) \: = \: \left( \frac{\partial}{\partial r} -
  \frac{l}{r} \right)
  B_{ilm}(r) \: .
\label{eq:deldef}
\end{eqnarray}
for the radial parts $V^{(a)}_{ilm}$ and $B^{(a)}_{ilm}$
due to $V_i({\bf r})$ and ${\bf B}_i({\bf r})$, respectively,
and
\begin{eqnarray}
W^{(1)}_{lm} (\hat{\bf r}) &=& \sum^{m'=1}_{m'=-1} Y^{m'}_{1}
(\widehat{{\bf e}_{A_0}})~\chi^{(1)}_{lmm'}~Y^{m+m'}_{l-1}
(\hat{\bf r}) \: ,
\nonumber \\
W^{(2)}_{lm} (\hat{\bf r}) &=& \sum^{m'=1}_{m'=-1} Y^{m'}_{1}
(\widehat{{\bf e}_{A_0}})~\chi^{(2)}_{lmm'}~Y^{m+m'}_{l+1}
(\hat{\bf r})~,
\end{eqnarray}
with
\begin{eqnarray}
\chi^{(1)}_{lmm'}
&=& \sqrt{\frac{3}{4\pi}} (-)^{m'} \sqrt{\frac{(l-m)
(l+m)}{(2l-1)(2l+1)}} \sqrt{\frac{2l-m'(m+m'(l+1))}{2(l+m'm)}} \: ,
\nonumber \\
\chi^{(2)}_{lmm'}
&=& \sqrt{\frac{3}{4\pi}} \sqrt{\frac{(l+1-m)
(l+1+m)}{(2l+1)(2l+3)}} \sqrt{\frac{2(l+1)+m'(m-m'l)}{2(l+1-m'm)}}~.
\end{eqnarray}
for the angular parts $W_{lm}^{(a)}$ ($a=1,2$). ${\bf e}_{A_0}$ is
the unit vector in the direction of the vector potential ${\bf A}_0$.

\section{The transition-matrix elements}
\label{sec:tme}

Now we are in the position to calculate the transition-matrix
elements
\begin{equation}
M_{n} ({\epsilon}_{f},{\bf k}_{\|}) =
\langle {\epsilon}_{f},{\bf k}_{\|}| G^{+}_{2} \Delta |
 n , {\bf q}_{\|} \rangle =
\int d^3r ~ \Psi^{(f)}_{{\bf k}_{\|}}({\bf r})^\dagger
\Delta({\bf r})
\Psi^{(n)}_{{\bf q}_{\|}}({\bf r}) \: .
\label{eq:matrixe}
\end{equation}
Inserting the one-center expansions for the initial and the final
states (\ref{eq:infinal}) and (\ref{eq:finstat}), respectively,
and using the decomposition (\ref{eq:deldec}) of the dipole
operator yields:
\begin{eqnarray}
   M_{n} ({\epsilon}_{f},{\bf k}_{\|}) & = & \sum_i
   \sum_{\kappa\mu\kappa'\mu'}
   A_{i\kappa\mu}(\epsilon_f, {\bf k}_\|)~
   M_{i\kappa\mu\kappa'\mu'}
     ({\epsilon}_{f},\epsilon_n({\bf q}_{\|}))~
   A_{i\kappa'\mu'}(\epsilon_{n}({\bf q}_\|),{\bf q}_\|)
   \nonumber \\
   & = &
   N_\| \sum_{i_\perp {i_{\sf A}}}
   \sum_{\kappa\mu\kappa'\mu'}
   A_{i_\perp{i_{\sf A}}\kappa\mu}(\epsilon_f, {\bf k}_\|)~
   M_{i_\perp{i_{\sf A}}\kappa\mu\kappa'\mu'}
       ({\epsilon}_{f},\epsilon_n({\bf q}_{\|}))~
   A_{i_\perp{i_{\sf A}}\kappa'\mu'}
       (\epsilon_{n}({\bf q}_\|),{\bf q}_\|)
   \: ,
   \nonumber \\
\label{eq:lattme}
\end{eqnarray}
where $N_\|$ is the number of two-dimensional unit cells per layer
and
\begin{equation}
   M_{i\kappa\mu\kappa'\mu'}(E_f,E_i) =
   \int_{\Omega_i} d^3r ~
   \Phi_{i\kappa\mu}^{(T)}(E_f,{\bf r})^\dagger
   \Delta_i({\bf r})
   \Phi_{i\kappa'\mu'}(E_i,{\bf r})
   \: ,
\label{eq:atomme}
\end{equation}
is the atomic matrix element of the dipole operator between the
phase functions at the respective eigenenergies. The integration
extends over the volume of the polyhedron $\Omega_i$. Since
$V_i({\bf r}) = {\bf B}_i({\bf r}) = 0$ and thus
$\Delta_i({\bf r}) = 0$ in the segments between $\Omega_i$ and the
bounding sphere $S_i$, the integration may be performed over $S_i$
as well which is more convenient. In the following we suppress the
atomic index $i$. Using (\ref{eq:delta1}) we get eight dipole matrix
elements,
\begin{equation}
   M_{\kappa\mu\kappa'\mu'}(E_f,E_i) =
   E_{fi} \sum_{dd'}^{\{ul\}}
   \int_{S} d^3r ~
   \Phi_{\kappa\mu}^{d(T)}(E_f,{\bf r})^\dagger
   \left(
   \Delta_V^{(dd')}({\bf r}) + \Delta_B^{(dd')}({\bf r})
   \right)
   \Phi_{\kappa'\mu'}^{d'}(E_i,{\bf r}) \: ,
\end{equation}
each of which can be separated into radial and angular parts:
\begin{eqnarray}
  M_{\kappa\mu\kappa'\mu'}(E_f,E_i) ~=~ E_{fi}
  \sum_{\kappa'' \mu'' \kappa''' \mu'''}
  \sum_{lm} \Bigg\{ \hspace{8mm} &&
\nonumber \\
\sum_{a=1}^2~
    {\cal R}_I^{(uu)}
        \Big[
    V_{lm}^{(a)}
        \Big]~
    {\cal A}^{\mu''\mu'''}_{\kappa''\kappa'''}
        \Big[
    W^{(a)}_{lm}
        \Big]
&-& \frac{\omega}{c}~
    {\cal R}_I^{(ul)}
        \Big[
    V_{lm}
        \Big]~
    {\cal A}^{\mu''\mu'''}_{\kappa''-\kappa'''}
        \Big[
    \mbox{\boldmath $\sigma$} {\bf A}_0
    Y^{m}_{l}
        \Big]
\nonumber \\
- ~ \frac{\omega}{c}~
    {\cal R}_I^{(lu)}
        \Big[
    V_{lm}
        \Big]~
    {\cal A}^{\mu''\mu'''}_{-\kappa''\kappa'''}
        \Big[
    \mbox{\boldmath $\sigma$} {\bf A}_0
    {Y^{m}_{l}}
        \Big]
&-& \sum_{a=1}^2~
    {\cal R}_I^{(ll)}
        \Big[
    V_{lm}^{(a)}
        \Big]~
    {\cal A}^{\mu''\mu'''}_{-\kappa''-\kappa'''}
        \Big[
    W^{(a)}_{lm}
        \Big]
\nonumber \\
+ ~ \sum_{a=1}^2~
    {\cal R}_I^{(uu)}
        \Big[
    B_{lm}^{(a)}
        \Big]~
    {\cal A}^{\mu''\mu'''}_{\kappa''\kappa'''}
        \Big[
    \mbox{\boldmath $\sigma$} {\bf b}
    W^{(a)}_{lm}
        \Big]
&-& \frac{\omega}{c}~
    {\cal R}_I^{(uu)}
        \Big[
    B_{lm}
        \Big]~
    {\cal A}^{\mu''\mu'''}_{\kappa''\kappa'''}
        \Big[
    \mbox{\boldmath $\sigma$} {\bf A}_0 \times {\bf b}
    Y^{m}_{l}
        \Big]
\nonumber \\
+ ~ \sum_{a=1}^2~
    {\cal R}_I^{(ll)}
        \Big[
    B_{lm}^{(a)}
        \Big]~
    {\cal A}^{\mu''\mu'''}_{-\kappa''-\kappa'''}
        \Big[
    \mbox{\boldmath $\sigma$} {\bf b}
    W^{(a)}_{lm}
        \Big]
&-& \frac{\omega}{c}~
    {\cal R}_I^{(ll)}
        \Big[
    B_{lm}
        \Big]~
    {\cal A}^{\mu''\mu'''}_{-\kappa''-\kappa'''}
        \Big[
    \mbox{\boldmath $\sigma$} {\bf A}_0 \times {\bf b}
    Y^{m}_{l}
        \Big]
\Bigg\}
\nonumber \\ &&
\label{eq:me}
\end{eqnarray}
with $I \equiv (\kappa\mu\kappa'\mu'\kappa''\mu''\kappa'''\mu''')$
and:
\begin{eqnarray}
  {\cal R}_I^{(dd')} \Big[ f \Big] & \equiv &
  \int_0^R r^2 dr~
  \phi^{d}_{\kappa''\mu''\kappa\mu}(r,E_f)
  ~f(r)~
  \phi^{d'}_{\kappa'''\mu'''\kappa'\mu'}(r,E_i) \: ,
\label{eq:rme}
\\
  {\cal A}_{\kappa\kappa'}^{\mu\mu'} \Big[ f \Big] & \equiv &
  \int_{(4\pi)} d\hat{\bf r}~
  \chi_{\kappa}^{\mu (T)}(\hat{\bf r})^\dagger
  ~f(\hat{\bf r})~
  \chi_{\kappa'}^{\mu'}(\hat{\bf r} ) \: ,
\label{eq:ame}
\end{eqnarray}
for $d=u,l$ and $d'=u,l$.

This completes our formalism. From the different contributions
(\ref{eq:me}) we can calculate the atomic matrix element
(\ref{eq:atomme}) for each layer $i_\perp$ and each atom
${i_{\sf A}}$ in the two-dimensional unit cell.
Inserting into (\ref{eq:lattme}),
one obtains the full transition-matrix element (\ref{eq:tmedef})
which enters the expression (\ref{eq:iphototme}) for the PES
intensity. For a single point $(\epsilon_f, {\bf k}_\|)$ in the
spectrum, it is necessary to determine the matrix elements
at the energies $E_f=\epsilon_f$ and $E_i=\epsilon_n({\bf q}_\|)$
for each $n$ (${\bf q}_\|$ is fixed) where there is a nonnegligible
contribution of the imaginary part of the initial-state Green
function $G_{nn'}(E_1,{\bf q}_\|)$.

The different types of radial matrix elements (\ref{eq:rme})
result from the different possible combinations of the upper and
lower components of the Dirac spinor. In the nonrelativistic case
only the $d=d'=u$ term survives. All one-dimensional integrals are
well defined since for $r\mapsto 0$ there is a $r^{-2}$
singularity of $f(r)$ at most. This can be seen from Eq.\
(\ref{eq:deldef}). Furthermore, the radial parts of the phase
functions are regular at $r=0$ [Eq.\ (\ref{eq:diracatom1})].

For the paramagnetic case ${\bf b}=0$ there are two types of
angular matrix elements (\ref{eq:ame}). The first in Eq.\
(\ref{eq:me}), $\sim W_{lm}^{(a)}$, is well known from the
nonrelativistic theory \cite{gbb}. For spherically symmetric
potentials $V_i({\bf r})=V_{i00}(r) Y_0^0(\hat{\bf r})$
[$l=m=0$ in (\ref{eq:me})] only the
term for $a=2$ yields a nonvanishing contribution. In this
case the angular matrix element further reduces to the muffin-tin
form \cite{PLNB97}. The second type of matrix elements,
$\sim \mbox{\boldmath $\sigma$} {\bf A}_0 Y^{m}_{l}$,
represents a relativistic correction \cite{gbb}.
In the ferromagnetic case two new types of matrix elements,
$\sim \mbox{\boldmath $\sigma$} {\bf b} W^{(a)}_{lm}$,
and
$\sim \mbox{\boldmath $\sigma$} {\bf A}_0 \times {\bf b} Y^{m}_{l}$,
occur.
All angular matrix elements are simply related to usual
Gaunt coefficients and can thus be calculated analytically.

\section{Summary}
\label{sec:con}

In this paper we have described in detail a new formalism to
evaluate Pendry's formula for the one-step model of (inverse)
photoemission spectroscopy. The initial-state Green function
represents the central physical quantity of the one-step model.
Compared with previous work, we propose a different treatment of
the initial state and determine the Green function from the Dyson
equation rather than within the DFT-LDA ground-state theory.
Generally, this is a necessary condition for a correct description
of one-particle excitations. Given the electronic self-energy, our
formalism allows to include the corresponding many-body effects
such as temperature dependencies, quasi-particle damping, band
narrowing, satellites, etc.\ in the one-step description of the
photoemission process. Within the framework of the original theory
\cite{Pen74,Pen76,HPT80} and its subsequent improvements
\cite{dev}, one is unable to deal with an in general {\em
nonlocal} self-energy correction since this is incompatible with
the KKR multiple-scattering formalism on which the original theory
is solely based \cite{PLNB97}. Our main idea is thus to
disentangle the calculation of the Green function from the
calculation of the dipole matrix elements. This is achieved by
considering the low-energy LDA eigenstates merely as basis states
to set up the Dyson equation. Solving the Dyson equation to get
the PES raw spectrum is then independent from the calculation of
the actual dipole matrix element. Note that it is necessary to
work within an {\em eigen}state basis of the LDA Hamiltonian since
this leads via Eq.\ (\ref{eq:diprel}) to the important spatial
decomposition (\ref{eq:deldec}) of the dipole operator which
eventually allows for a convenient calculation of the
transition-matrix elements.

The present paper has shown that the alternative evaluation of
the one-step model can formally be performed consistently for
very general cases, namely for complex geometries with more than
one atom per unit cell, for general space-filling nonspherical
LDA potentials, for situations which require a fully relativistic
treatment, and for the ferromagnetic case, in particular when
exchange and spin-orbit splitting must be treated on equal footing.

It goes without saying that the presented formalism can also be
applied without taking into account the self-energy correction.
Even in the case $\Sigma \equiv 0$ it should be advantageous since
it is more transparent and more simple compared with the
corresponding (full-potential, relativistic) theory \cite{gbb}
based on the original concept \cite{Pen76}: (i) Simple
one-dimensional integrals (\ref{eq:rme}) have to be performed
instead of two-dimensional integrations \cite{gbb} for the
calculation of the radial matrix elements. Furthermore, their
number is reduced considerably. (ii) Only the regular instead of
both, the regular and the irregular solution of the atomic Dirac
equation is needed [see Eq.\ (\ref{eq:diracatom1})]. (iii) There
are no additional intra-layer and inter-layer contributions as in
the original theory \cite{gbb}. The partitioning of the
photocurrent into atomic, intra-layer and inter-layer
contributions in the original formulation is only due to formal
reasons, and a direct physical interpretation is difficult. (iv)
Contrary, the separate calculation of the initial-state Green
function and the matrix elements allows to distinguish clearly
between the raw spectrum and its modifications due to secondary
effects.

The ferromagnetic Gd(0001) surface represents a prototype system
for the application of the theory. Since $Z=64$ for Gd, a fully
relativistic evaluation of the one-step model is meaningful even
if it is based on a potential and self-energy input from
scalar-relativistic LDA calculations \cite{Nolgd} and from
two-component many-body theory \cite{Nolgd}, respectively. Recent
many-body calculations show that striking temperature-dependent
correlation effects should be observable in the (5d, 6s)
conduction band as a consequence of intra-atomic exchange coupling
to the subsystem of localized and ferromagnetically ordered 4f
moments \cite{Nolgd,DDN}. The interpretation of several PES/IPE
spectra from Gd (cf.\ e.~g.\ Ref.\ \cite{DDN}) is controversial up
to now and may be resolved by supplementing the
electronic-structure calculations with a reliable treatment of the
secondary effects. Calculations based on the presented formulation
of the one-step model are intended for the future.

\section*{Acknowledgements}
This work was supported in part by the BMBF
within the Verbundprojekt
``Elektronische Struktur und Photoemission von hochkorrelierten
intermetallischen seltenen Erdverbindungen''
(contract no.: 05605MPA0)
and in part by the Deutsche Forschungsgemeinschaft within the
Sonderforschungsbereich 290 (``Metallische d\"unne Filme: Struktur,
Magnetismus und elektronische Eigenschaften'').

\end{document}